\documentclass[10pt,english,a4paper]{article}
\usepackage{amssymb}
\usepackage{amsmath}
\usepackage{babel}
\usepackage{multicol}
\usepackage{epsfig}

\begin{document}
{\Large The simplest derivation of the Lorentz transformation}\\

{\footnotesize \bf J.-M. L\'evy \\ 
\it Laboratoire de Physique Nucl\'eaire et de Hautes Energies,
CNRS - IN2P3 - Universit\'es Paris VI et Paris VII, Paris.  \\ \it Email: jmlevy@in2p3.fr}\\
\begin{abstract}
The Lorentz transformation is derived from the simplest thought experiment by using the simplest vector 
formula from elementary geometry. The result is further used to obtain general velocity and acceleration 
transformation equations.
\end{abstract}

\hspace{1cm}{\it Submited to the American Journal of Physics}
\begin{multicols}{2}
\section{Introduction}
Many introductory courses to special relativity (SR) use thought experiments in order to demonstrate time dilation 
and length contraction from the two Einstein's postulates by using conceptual devices like the well known light clock
or variants thereof (see below or e.g. \cite{Krane};  \cite{Mathews} gives an extensive bibliography)
However, once these two effects are established, most authors start again from the postulates to derive the Lorentz 
transformation (LT), taking the route which is usual in advanced texts but which is certainly not the easiest one 
to begin with. \\
However, deriving the LT directly from these effects is possible and has obvious advantages for beginners in 
terms of simplicity. It allows, for example, to  bypass the use of the requirements of group structure and 
linearity. Paramount though they are in fundamental physics, dispensing with these considerations allows for a 
very direct first contact with the conceptually demanding subject of SR \cite{Levy}. More elaborate derivations
from fundamental principles can be left for a second pass.\\

In the present article, we show that the LT can be derived from length contraction through a purely geometrical 
argument which amounts to expressing the basic vector addition formula in the two frames at stake successively.
 This reasoning leads to a very simple and possibly new way of writing the space part of the LT, which in turn 
allows an easy derivation of the velocity and acceleration tranformations as well.\\
A paper in which our type of exposition was used already appeared in the American Journal of Physics a long time ago 
\cite{Park}. However, the author of \cite{Park} missed what we think is the easiest way to derive the time 
transformation formula and was led to obtain it through a rather contrived argument, introducing an artificial 
extension of the 'time interval'. Also, as in most papers on the subject, the derivation was limited to transformations 
between two reference frames in the so-called 'standard configuration' \cite{Rindler}, viz. parallel axes, $OX'$ 
sliding along $OX$ with co\"{\i}ncident space-time origins.\\

This paper is organised as follows: in order to prevent possible objections which are often not taken care of in the 
derivation of the two basic effects using the light clock, we start by reviewing it briefly in 
section 2. The LT between two frames in 'standard configuration' is first derived from length contraction in 
section 3. Section 4 deals with the more general case of an arbitrarily oriented relative velocity. Using the 
expression obtained here, the velocity and acceleration transformations are found in Section 5. Section 6 contains 
our summary and conclusions.\\  

\section{Time dilation and length contraction}
\subsection{The light clock}
The light clock is a simple conceptual device to demonstrate the basic effects of SR starting from 
Einstein's two postulates, viz. the principle of relativity and the invariance of the velocity of light in a change 
of inertial frame. \\
 We imagine (see fig. 1) a light signal bouncing back and forth between two
parallel mirrors maintained a constant distance apart with the aid of pegs. The signal triggers the registering of a 
tick each time it hits what we define as the 'lower' mirror (fig.1 a). The question of how this device can be 
practically constructed does not concern us. We simply assume that there is a way to sample the signal in order to
produce the tick and to compensate for the loss of light incured thereof. We thus have a kind of perfect clock the 
period of which is $T_0 = \frac{2L_0}{c}$ with $L_0$ the distance between the mirrors and $c$ the speed of 
light. \\
\subsubsection{Time dilation}
Let's now look at the clock in a frame wherein it travels at constant speed $v$ in a direction parallel to the planes 
of the mirrors. We might assume that the mirrors are constrained to slide in two parallel
straight grooves which have been engraved a constant distance $L_0$ apart, previous to the experiment, so that there
can't be any arguing about a variation of the pegs length when they are moving.\\ By the first postulate, this
moving clock must have the same period in its rest frame than its twin at rest in the laboratory.\\
On the other hand, it is obvious that the length traveled by the signal in the lab is longer than the 
length it travels in the clock rest frame (see fig.1 b.) If $T$ is the interval between two ticks in the lab, then by 
Einstein's second postulate and Pythagora's theorem we have that $(c T/2)^2 = L_0^2 + (v T/2)^2$ from which $T = 
\frac{T_0}{\sqrt{1-(\frac{v}{c})^2}}$ follows, which shows that the moving clock runs more slowly in the lab than 
its still twin.
\subsubsection{Length contraction} We now imagine that the moving clock is traveling in a direction perpendicular to 
the plane of its mirrors relative to the lab observer. In this case, no check can be kept of the inter-mirror 
distance. 
To make sure that (for the same $v$) the clock period hasn't changed, we can imagine it accompanied by an identical 
second clock oriented as before with respect to its lab velocity. 
Both clocks having the same period in their 
common rest frame and the clock moving parallel to its mirrors having period $T$ in the lab, we can be sure that the 
first clock period as measured in the lab frame hasn't changed either. Anticipating the result which will be 
forced upon us, we call $L$ the inter-mirror distance as measured in the laboratory frame.
Consider now the time taken by the light signal to make its two-way travel in this later frame; we see that it
needs $\frac{L}{c+v}$ for the lower mirror to upper mirror part and $\frac{L}{c-v}$ for the return part.
Since the total must equal $T$, one is forced to conclude that $L = L_0\sqrt{1-(\frac{v}{c})^2}$
 \\
That the distances in the directions orthogonal to the motion are not changed can be demonstrated by invoking grooves 
arguments like the one we used for the time dilation derivation.  
\psfig{file=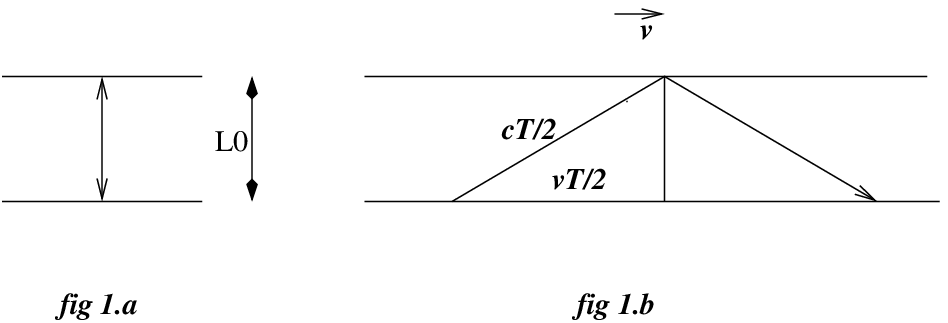}
\begin{center}
The light clock at rest (left) and moving(right)
\end{center}
\section{Lorentz transformation along the $x$ axis}
Let us now envision two frames in 'standard configuration' with $K'$ having velocity $\bf v$ with respect to $K$ and 
let
$x,t$ (resp. $x',t'$) be the coordinates of event $M$ in the two frames. Let $O$ and $O'$ be the spatial origins of
the frames; $O$ and $O'$  co\"{\i}ncide at time $t = t' = 0$\\

Here comes the pretty argument: all we have to do is to express the relation ${\bf OM = OO' + O'M}$ between 
vectors (which here reduce to oriented segments) in both frames.\\
In $K$, $\overline{OM} = x$, $\overline{OO'} = vt$ and $\overline{O'M}$ {\it seen from $K$} is $\frac{x'}{\gamma}$ 
with $\gamma = \frac{1}{\sqrt{1-(\frac{v}{c})^2}}$ since $x'$ is $\overline{O'M}$ as measured in $K'$
Hence the first relation: $$x = vt + \frac{x'}{\gamma}$$ 
In $K'$, $\overline{OM} = \frac{x}{\gamma}$ since $x$ is $\overline{OM}$ {\it as measured in $K$}, $\overline{OO'} = 
vt'$ 
and $\overline{O'M} = x'$. Hence a second relation: $$\frac{x}{\gamma} = vt' + x'$$ 
The first relation yields immediately $$x' = \gamma(x-vt)$$ which is the 'space' part of the LT 
and the second relation yields the inverse $$x = \gamma(x'+vt')$$ of this 'space part'. Eliminating $x'$ between 
these two leads quickly to the formula for the transformed time: $$t' = \gamma(t-vx/c^2)$$ the inverse of which 
could 
easily be found by a similar elimination of $x$.\\
Needless to say, coordinates on the $y$ and $z$ axes are unchanged for the already stated reason that distances do 
not vary in the directions perpendicular to the velocity. The contraction
is therefore limited to that part of the coordinate vector which is parallel to the relative velocity.
\section{The case of an arbitrary velocity}
In the following, $\bf v$ will denote the velocity vector of $K'$ w.r.t. $K$ and
$\bf r$ (resp. $\bf r'$) the position vector of the event under consideration as measured in frame $K$ (resp $K'$). 
We further define $\bf u = \frac{v}{|v|}$ the unit vector parallel to $\bf v$. \\
From our findings of section 2, we see that only the component of $\bf r$ parallel to $\bf v$ is affected when 
looking at it from the other frame, whilst the normal components are unchanged. We resolve $\bf r$ into parallel and 
perpendicular components according to $\bf r = u u.r +(1 - u\otimes u)r = r_{\parallel} + r_{\perp}$  where the dot 
stands for the 3-space scalar product, $\bf 1$ is the identity operator and $\bf u\otimes u$ is the dyadic which 
projects out the component parallel to $\bf u$ from the vector it operates upon, viz $\bf (u\otimes u)V = (u.V) u$. 
\\ The operator which 
contracts the projection on $\bf u$ by $\gamma$ whilst leaving the orthogonal components unchanged must yield: $\bf u 
\frac{u.r}{\gamma} + (1 -u\otimes u)r = (1 +\frac{{\rm 1}-\gamma}{\gamma}u\otimes u)r$. Let us therefore define $\bf 
Op(\gamma^{\rm -1}) = 1 + \frac{{\rm 1}-\gamma}{\gamma} u \otimes u$ 
The inverse operator  must correspond to multiplication of the longitudinal part by $\gamma$ and is therefore
$\bf Op(\gamma) = Op(\gamma^{\rm -1})^{\rm -1} = 1 + (\gamma-{\rm 1}) u\otimes u$ as can also be checked by 
multiplication. Note that these operators are even in $\bf u$ and therefore independent of the orientation of 
$\bf v$.\\

Mimicking what has been done in section 3, let us 
now write $\bf OM = OO' + O'M$ (these are vectors now, no longer oriented segments) taking care of the invariance of 
the orthogonal parts. We get in frame $K$: $$\bf r = v{\rm t} + Op(\gamma^{\rm -1})r'$$ and in frame $K'$: 
$$\bf Op(\gamma^{\rm -1})r = v{\rm t'} + r'$$
 
Using the inverse operator the first relation yields immediately: 
$$\bf r' = Op(\gamma)(r - v{\rm t}) = (1 + (\gamma-{\rm 1}) u\otimes u)(r - v{\rm t})$$ which is probably the 
simplest way 
to write the space part of the rotation free homogenous LT. The usual $\gamma$ factor of the 
one dimensionnal transformation is simply replaced by the operator $\bf Op(\gamma)$\\
By feeding this result into the second relation above, we find:
$$\bf Op(\gamma^{\rm -1})r = v{\rm t'} + Op(\gamma)(r-v{\rm t})$$ or, using $\bf Op(\gamma)v = \gamma v$ and 
with the explicit form of $\bf Op$:
$$(\frac{1-\gamma}{\gamma} -(\gamma-1))\frac{\bf v v.r}{v^2} +\gamma {\bf v} {\rm t} = {\bf v}{\rm t'}$$ which, using 
$1-\gamma^2 
= 
-(\frac{v}{c})^2\gamma^2$ and crossing away $\bf v$ on both sides yields:
$$t' = \gamma (t-\frac{\bf v.r}{c^2})$$ i.e. the time transformation equation. 
\section{Velocity and acceleration transformations}

\subsection{Velocity}
The two formulas thus obtained for the L.T. are so simple that they can readily be used to yield the
velocity transformation equation without the need of complicated thought experiments and algebraic manipulations.
Differentiating $\bf r'$ and $t'$ w.r.t. $t$ and taking the quotient of the equalities thus obtained yields
(with ${\bf V'} = \frac{d{\bf r'}}{dt'}$ and ${\bf V} = \frac{d{\bf r}}{dt}$) $$\bf V' = \frac{\rm 
1}{\gamma}\frac{({\rm 1} + (\gamma-{\rm 1}) u\otimes u)(V - v)}{{\rm 1}-\frac{v.V}{\rm c^2}}$$ which is the general 
velocity transformation formula. 
\subsection{Acceleration}
Using the compact $\bf Op$ notation helps to keep things tidy when differentiating $\bf V'$ w.r.t. $t$; dividing by 
the derivative of $t'$ one finds $$\bf A' = \frac{\rm 1}{\gamma^2}\frac{Op(\gamma) A ({\rm 1}-\frac{v.V}{\rm c^2}) 
+Op(\gamma)(V-v)\frac{v.A}{\rm c^2}}{({\rm 1}-\frac{V.v}{\rm c^2})^{\rm 3}}$$
Expliciting $\bf Op$, simplifying and regrouping terms, one obtains after a page of algebra: $$\bf A' = 
\frac{A-\frac{\gamma}{\gamma+{\rm 1}}\frac{v.A v}{\rm c^2}+\frac{v \wedge(V \wedge A)}{\rm c^2} }{\gamma^{\rm 2} ({\rm 1}-\frac{V.v}{\rm c^2})^{\rm 3}}$$
By making the necessary substitutions: $\bf V \rightarrow u'$, $\bf V' \rightarrow u$, $\bf v \rightarrow -V$ and 
specializing to $\bf V$ parallel to $Ox$, one can easily check that the component equations derived from this  
general formula agree with those published in \cite{Mathews}. They have been, however, derived with much less 
effort.\\
As an example of use of this acceleration transformation, by specializing to $\bf V=v$ and $\bf v.A = 
0$, one gets $\bf A' = \gamma^{\rm 2} A$ retrieving the known result that a particle in a circular storage ring undergoes
a proper ($\bf A'$) acceleration that is a factor $\gamma^2$ larger than the lab ($\bf A$) acceleration. Moreover, the 
two accelerations are parallel, which is far from obvious a priori. Observe in this respect, that all the terms which 
can make $\bf A'$ and $\bf A$ different in direction as well as in size vanish in the $c \rightarrow \infty$ limit, 
consistent 
with the fact that acceleration is an invariant quantity under a change of inertial frame in newtonian physics.  \\
Setting $\bf V = v$ and taking $\bf v$ parallel to $\bf A$ we also retrieve another known fact: a particle in
rectilinear motion undergoes a proper acceleration which is larger than its lab acceleration by a factor $\gamma^3$. 
\section{Summary and conclusion}
We have shown that the general rotation free homogenous LT can be derived once length contraction 
has been established by writing the elementary vector relation (sometimes dubbed 'Chasles' relation) ${\bf OM = OO' + 
O'M}$ in the two frames considered \cite{MacDo}. The extension from the special one dimensional case to the 
3-dimensional case
is completely straightforward. The formula thus obtained allows for a simple derivation of the velocity and 
acceleration transformations without the need for complicated thought experiments and algebraic manipulations 
beyond what college students are used to.
\begin{footnotesize}

\end{footnotesize}

\end{multicols}


\begin{thebibliography}{2}
\bibitem{Levy}For the sake of completeness, it must be mentionned that the LT can in fact be derived by using 
the sole relativity principle (first Einsein postulate) and dispensing with the second (invariance of the speed of 
light)\cite{Rindler2} This is certainly most appealling from a puristical viewpoint, but it demands more abstract 
work which is what we are trying to avoid here, having in mind beginning students.\
\end{thebibliography}
\end{document}